\documentclass[11pt,a4paper]{article}
\usepackage[dvips]{graphicx}
\newlength{\extraspace}
\newlength{\extraspaces}
\newcommand{\be}{\begin{equation}
\addtolength{\abovedisplayskip}{\extraspaces}
\addtolength{\belowdisplayskip}{\extraspaces}
\addtolength{\abovedisplayshortskip}{\extraspace}
\addtolength{\belowdisplayshortskip}{\extraspace}}
\newcommand{\ee}{\end{equation}}

\newcommand{\ba}{\begin{eqnarray}
\addtolength{\abovedisplayskip}{\extraspaces}
\addtolength{\belowdisplayskip}{\extraspaces}
\addtolength{\abovedisplayshortskip}{\extraspace}
\addtolength{\belowdisplayshortskip}{\extraspace}}
\newcommand{\ea}{\end{eqnarray}}

\newcommand{\newsection}[1]{
\vspace{15mm}
\pagebreak[3]
\addtocounter{section}{1}
\setcounter{equation}{0}
\setcounter{subsection}{0}
\setcounter{footnote}{0}
\begin{flushleft}
{\large\bf \thesection. #1}
\end{flushleft}
\nopagebreak
\medskip
\nopagebreak}

\newcommand{\uf}\uparrow
\newcommand{\df}\downarrow
\newcommand{\Td}\bigtriangledown

\begin{document}

\addtolength{\baselineskip}{.8mm}

\thispagestyle{empty}

\begin{flushright}
{\sc GEF-Th}-10/1997\\
{\sc hep-th}/97\\
September 1997
\end{flushright}
\vspace{.3cm}

\begin{center}
{\Large\bf{On the ERG approach in $3 - d$ Well Developed
Turbulence\footnote{Partially supported by MURST, Italy}}}\\[12mm]

{\bf Renzo Collina\footnote{collina@grp4.ge.infn.it} 
and Paolo Tomassini\footnote{tomassini@infn.ge.infn.it}}\\[2mm]

{\it Dipartimento di Fisica dell'Universit\`a Genova,}\\[2mm]
{\it Istituto Nazionale di Fisica Nucleare, Sezione di Genova,}\\[2mm]
{\it via Dodecaneso, 33, 16146 Genova, Italy}\\[16mm]

{\bf Abstract}
\end{center}

\noindent
We apply the method of the Renormalization Group (GR), following the
Polchinski point of view, to a model of well developed and
isotropic fluid turbulence. 
The Galilei-invariance is preserved and a universality
behavior, related to the change of the stochastic stirring force, is evident by
the numerical results in the inertial region, where a scale-invariant behavior
also appear. The expected power law of the energy spectrum ($q^{-{5\over 3}}$)
is obtained and the Kolmogorov constant $C_K$ agrees with experimental data.

\vfill
\newpage

\newsection{Introduction}

\noindent
The incompressible fluid turbulence, at very high Reynolds number (${\cal{R}}$),
is described  by physical systems with a large number of degrees of freedom that
are  characterized by a substantial scale invariance. These peculiarities are
investigated in Statistical Hydrodynamics by  models in which the
Navier-Stokes equations are supplemented  with stochastic force terms and the
analogies between scaling at high-Reynolds-number turbulence regime and
equilibrium critical 
phenomena have been proposed. Attempts have been made to
apply the powerful {\it renormalization-group} (RG) 
methods, which have proved so
effective and illuminating in critical phenomena, to turbulence scaling
\cite{FSN},\cite{DM},\cite{YO}. De Dominicis and Martin \cite{DM} analysed these
models near the critical dimension with a perturbative approach.  Yakhot and
Orszag \cite{YO} 
have extrapolated the perturbative RG study from the critical to
physical dimension and, on this basis they predicted, in a long inertial range,
various scaling laws, such as the Kolmogorov energy spectrum \cite{KO},
including constant factors as the Kolmogorov constant, the
Obukhov-Corrsin constant and the turbulent Prandtl number. But the numerical
success of the Yakhot and Orszag's theory does not rest on firm theoretical
grounds. Indeed the 
critical dimension ($d=7$) is  very far from the physical one
($d=3$) and it is not evident that an $\epsilon$ expansion is the best procedure
\cite{GE}. Another difficulty of the Yakhot and Orszag's method is related to
the breaking of the Galilei invariance produced by the Wilson-Kadanoff RG
analysis in the presence of an infinite number of marginal diagrams \cite{GE}.
For these reasons the perturbative RG approach requires some ad hoc hypothesis
\cite{YO} and produces only a phenomenological
description of turbulence \cite{GE},\cite{GE96},\cite{Com}.

The goal of this paper is that of reconsidering the RG approach for the
homogeneous isotropic turbulence, for an incompressible fluid, at very high
Reynolds number, but in a non perturbative context. The point of view is very
similar to that introduced by Polchinski \cite{J.Pol}, \cite{GG}, \cite{CB} 
in the
context of the Quantum Field Theory (QFT). The analogy with QFT follows from
the description of our model in terms of a path-integral representation of the
probability generating functionals, using a so-called Martin-Siggia-Rosen (MSR)
action \cite{MSR}. 

The dynamics of the model is characterized by the presence of two
scales; one of them corresponds to the macroscopic size $L$ of the system, 
and the mechanism which maintains the energy stationary is operating at this
scale. The second, the Kolmogorov scale $\eta_d$, is an internal, 
typically smaller
scale, at which the energy, after its transfer by 
a cascade process \cite{Com}, is
dissipated. When the dimensionless quantity ${L\over \eta_d} = {\cal{R}}^{3\over
4}$ is very large the system admits a wide {\it inertial range}, where its
statistical properties are universal, homogeneous, isotropic and self-similar. 
As in the Yakhot and Orzag's paper \cite{YO},  the model describes true
turbulence only in this long inertial range and the manifest scale invariance of
this region is interpretable as the existence of a critical
phase in the infinite volume limit.

The analysis of this paper is centered on the study of the  flow equations for
the effective action of an infrared regularized theory. This method involves an
infinity of coupled differential equations and necessary requires an
approximation scheme. The perturbative approximation \cite{BDM}, for the
previous arguments, preserves an infinite number of terms and, in our case, no
summability criterion of the perturbative series is known. In other schemes, 
as the derivative expansion approximation \cite{TM3}, 
\cite{TM2} the correlations
functions are strongly dependent on the particular 
form of the stirring force also
in the long inertial region. The problem 
of the approximation has its solution in
the fact that, in the momentum space, the size of the spectrum of two point
stirring force correlation function is of 
order $O\left({1\over L}\right)$. From
this property, using the same flow equation, it is possible to see that, in the
inertial range (and only in this region), some proper correlation functions are
smaller then others.

The paper is organized as follow: 

In Section 2 we summarize the model concerning the Navier-Stokes equation with a
stochastic force term following the Martin, Siggia and Rose point of view.

In Section 3 we analyze the Galilei-invariance of the model and we write the
Ward Identities for the functional generators.

Section 4 is devoted to the equation flow referred to the scale parameters which
characterize the model; we also analyze the boundary conditions and some
properties of the vertex functions are discussed in Appendix A.

Section 5 contains a discussion concerning the approximation scheme with the
relevant hypotheses and a plausibility argument which further considered in
Appendix B.

Some numerical results, concerning the renormalized viscosity and the Kolmogorov
constant, are contained in Section 6.

Finally some conclusive remarks are contained in Section 7.

\newsection{The model}

\noindent
Let us consider the 
Navier-Stokes equations, supplemented with a stochastic force
term,
\be
\label{NS}
{\partial\over \partial t}u^{\alpha}(t, \vec{x}) + u^{\beta}(t, \vec{x})
{\partial\over \partial x^{\beta}}u^{\alpha}(t, \vec{x}) = - {1\over \rho}
{\partial\over \partial x_{\alpha}}p(t, \vec{x}) 
+ \nu\Td^2u^{\alpha}(t, \vec{x})
+ f^{\alpha}(t, \vec{x})
\ee
and with the incompressibility constraint
\be 
\label{incomp}
{\partial\over \partial x^{\alpha}}u^{\alpha}(t, \vec{x}) = 0.
\ee
The fields $u^{\alpha}$ are the three components of the velocity field and
$p$ is the pressure field; the effects of the boundary conditions over the 
zero mean value field $u^{\alpha}$ are summarized by the stochastic source
$f^{\alpha}$ which is a Gaussian random force with
mean zero and covariance
\be
\label{covar}
\langle f^{\alpha}(t, \vec{x})f^{\beta}(t^{\prime}, \vec{y})\rangle =
F^{\alpha\beta}\left(t, t^{\prime}; \vec{x}, \vec{y}\right) =
2P^{\alpha\beta}(\vec{\Td}_x)N(\vec{x}-\vec{y})\delta(t - t^{\prime}). 
\ee
$P^{\alpha\beta}(\vec{\Td}_x)$ is the projection onto solenoidal vector fields,
required to maintain the incompressibility constraint (\ref{incomp}). Since
the {\it stirring force} $f^{\alpha}$ should mimic the instabilities occurring
near the boundaries, its correlation length must be of the same order of the
system size. In other words if $L$ 
is the length corresponding to the size of the
system and
\be
\label{noise}
\langle f^{\alpha}(q^0, \vec{q})f^{\beta}(q^{0\prime}, \vec{q}^{\prime})\rangle 
= F^{\alpha\beta}(q^0, q^{0\prime}; \vec{q}, \vec{q}^{\prime}) =
2(2\pi)^4P^{\alpha\beta}(q)h(q)\delta^3(\vec{q} + \vec{q}^{\prime})\delta(q^0 +
q^{0\prime}).  
\ee
is the Fourier transform of (\ref{covar}), the function $h(q)$ must vanish
for $q>{1\over L} = m$. The explicit form of this function is not very important
if the self-similarity behaviour is realized, but in the stationary limit the
energy dissipated by the viscous forces must be compensated by the energy
introduced through the boundaries and a physical normalization condition is
needed, e. g. 
\be
\label{E}
\langle {\cal{E}}\rangle = 2\int {d^3q\over (2\pi)^3}h(q),
\ee
where ${\cal{E}}$ is the rate of power dissipated by a unit mass of fluid and
can be calculated by the local quantity \cite{Com},\cite{Lang}

$${\cal{E}}(t, \vec{x}) = {1\over 2}
\nu\left(\partial_{\alpha}u_{\beta}(t, \vec{x}) 
+ \partial_{\beta}u_{\alpha}(t, \vec{x})\right)^2.$$

\noindent
The probability generating functional is given by the functional integral
\be
\label{generA}
{\cal{W}}\left(J, \hat{J}, p, \hat{p}\right) = \int
{\cal{D}}u{\cal{D}}\hat{u}{\cal{D}}f e^{i\int d\hat{x} {\cal{L}}(u, \hat{u}, p,
\hat{p}, f, J, \hat{J})} e^{-{1\over 2}\int d\hat{x}
d\hat{y}f^{\alpha}(\hat{x})F^{-1}_{\alpha\beta}(\hat{x},
\hat{y})f^{\beta}(\hat{y})}  
\ee
where we have set $\hat{x}\equiv\left(t, \vec{x}\right)$. ${\cal{L}}$ is the
Navier-Stokes density of Lagrangian

$${\cal{L}}(\hat{x}) =\hat{u}^{\alpha}(\hat{x})\left({\partial\over
\partial t} - \nu\Td^2\right)u_{\alpha}(\hat{x}) + \hat{u}^{\alpha}(\hat{x})
u_{\beta}(\hat{x}){\partial\over \partial x^{\beta}}u_{\alpha}(\hat{x})$$ 
$$+ {1\over \rho}\hat{u}^{\alpha}(\hat{x}){\partial\over \partial x^{\alpha}}
p(\hat{x}) + {1\over \rho}\hat{p}(\hat{x}){\partial\over \partial x^{\alpha}}
u^{\alpha}(\hat{x}) + \hat{J}^{\alpha}(\hat{x})u_{\alpha}(\hat{x}) +
\hat{u}^{\alpha}(\hat{x})J_{\alpha}(\hat{x}).$$

\noindent
The field $\hat{u}^{\alpha}$ is the conjugate variable to the velocity field
$u^{\alpha}$, $p$ (the pressure field) and its conjugate variable $\hat{p}$ are
the two Lagrangian's multipliers related to the solenoidal constraint for the
fields $u^{\alpha}$ and $\hat{u}^{\alpha}$ and finally $J^{\alpha}$ and 
$\hat{J}^{\alpha}$ are external sources. In (\ref{generA}) it is possible
to integrate over the functional 
gaussian measure ${\cal{D}}f e^{-{1\over 2}\int 
f^{\alpha}F^{-1}_{\alpha\beta}f^{\beta}}$ and to reduce the space of the 
solutions only to 
the independent transverse degrees of freedom. We introduce the
transverse variables $v^{\alpha}=P^{\alpha\beta}(\Td)u^{\beta}$, 
$\hat{v}^{\alpha}=P^{\alpha\beta}(\Td)\hat{u}^{\beta}$ and, after the
integration over $f^{\alpha}$, we obtain \cite{GE96} the Martin, 
Siggia and Rose probability generating functional \cite{MSR}
\be
\label{MSRgener}
{\cal{W}}(J, \hat{J}) = \int{\cal{D}}v{\cal{D}}\hat{v}
e^{i\left(S(v,\hat{v}) 
+ \int d\hat{x}\hat{J}^{\alpha}(\hat{x})v_{\alpha}(\hat{x})
+ \int d\hat{x}\hat{v}^{\alpha}(\hat{x})J_{\alpha}(\hat{x})\right)}
\ee
where the action $S(v,\hat{v})$ is
\ba
&S(v,\hat{v}) = \int dt\int d^3x\hat{v}^{\alpha}(t,\vec{x})\left[
\left(\partial_t - \nu\Td^2\right)v_{\alpha}(t,\vec{x}) + v^{\beta}(t,\vec{x})
\partial_{\beta}v^{\alpha}(t,\vec{x})\right]\nonumber \\
&+ {i\over 2}\int dt\int d^3x\int d^3y\hat{v}^{\alpha}(t,\vec{x})N_{\alpha\beta}
(\vec{x}-\vec{y})\hat{v}^{\beta}(t,\vec{y}).
\label{MSRaction} 
\ea
The multitime and multipoint correlation functions are obtained by
functional derivatives of (\ref{MSRgener}) with respect to the external sources
$\hat{J}^{\alpha}$ and $J^{\alpha}$. The connected functional is defined as 
\be
\label{connesso}
ln{\cal{W}}(J, \hat{J}) = i{\cal{Z}}(J, \hat{J})
\ee
and the generic connected correlation function is given by
$$
\langle
v^{\alpha_1}(\hat{x}_1)...v^{\alpha_n}(\hat{x}_n)\hat{v}^{\alpha_{n+1}}
(\hat{x}_{n+1})...\hat{v}^{\alpha_{n+m}}(\hat{x}_{n+m})\rangle_c$$
\be
= \left.(-i)^{n+m}e^{-i{\cal{Z}}(J, \hat{J})}{\delta\over
\delta\hat{J}^{\alpha_1}(\hat{x}_1)} ...{\delta\over
\delta\hat{J}^{\alpha_n}(\hat{x}_n)} {\delta\over \delta
J^{\alpha_{n+1}}(\hat{x}_{n+1})}... {\delta\over \delta
J^{\alpha_{n+m}}(\hat{x}_{n+m})}e^{i{\cal{Z}}(J, \hat{J})}\right|_{\hat{J}=J=0} 
\ee
where the stability conditions
\be
\label{stability}
\left.{\delta{\cal{Z}}(J, \hat{J})\over 
\delta\hat{J}^{\alpha}(\hat{x})}\right|_{\hat{J}=J=0}
= \left.{\delta{\cal{Z}}(J, \hat{J})\over 
\delta J^{\alpha}(\hat{x})}\right|_{\hat{J}=J=0} = 0
\ee
are satisfied. Finally the convex vertex functional (the effective action)
$\Gamma(v, \hat{v})$ is defined by the Legendre transformation \cite{GE96}
\be
\label{vertex}
{\cal{Z}}(J, \hat{J}) = \Gamma(v, \hat{v}) + \int
d\hat{x}\left(\hat{J}^{\alpha}(\hat{x})v_{\alpha}(\hat{x}) 
+ \hat{v}^{\alpha}(\hat{x})J_{\alpha}(\hat{x})\right).
\ee
Since  from (\ref{vertex}) we have the relation

$${\delta\Gamma(v, \hat{v})\over \delta v^{\alpha}(\hat{x})} 
= - \hat{J}^{\alpha}(\hat{x}), \quad 
{\delta\Gamma(v, \hat{v})\over \delta \hat{v}^{\alpha}(\hat{x})} 
= - J^{\alpha}(\hat{x}),$$

\noindent
the stability conditions (\ref{stability}) are rewritten
\be
\label{prostab}
\left.{\delta\Gamma(v, \hat{v})\over \delta v^{\alpha}(\hat{x})}\right|_{u=
\hat{u}=0}  =  
\left.{\delta\Gamma(v, \hat{v})\over \delta \hat{v}^{\alpha}(\hat{x})}
\right|_{u=\hat{u}=0}  =  0.
\ee
We conclude this section observing that
the knowledge of the functional ${\cal{W}}(J, \hat{J})$, and then of
${\cal{Z}}(J, \hat{J})$ and $\Gamma(v, \hat{v})$, gives a complete solution of
the model in terms of correlation functions. The direct calculation of the
functional generator from the Navier-Stokes equation (\ref{NS}), with suitable
asymptotic boundary conditions, is a formidable task which may be 
simplified studying the constraints deriving from the symmetries of the model and
the scaling properties connected to the variation of the dimensional parameters.
The following sections are devoted to developing this point of view.

\newsection{Galilei invariance and Ward's identities}

\noindent
In the theory of turbulence we consider the very important hypothesis that, in
the limit of infinite Reynolds number, all the symmetries of the Navier-Stokes
equation, that are broken by the mechanism producing the turbulent flow, are
restored in a statistical sense at small scale and away from the boundaries
\cite{UF}.
In our model, the last term in eq.(\ref{MSRaction}) is Galilei invariant, as the
original Navier-Stokes equation, and this symmetry will be transferred to all
correlation functions. Then, as a physical system is described by its
symmetries, if scale invariance is also recovered in the
inertial range region, the model will provide a good  description of the 
isotropic
turbulence in this region. In this section we study the constraint deriving from
Galilei-invariance.

Let us consider the following coordinates transformations

$$t\rightarrow t^{\prime} = t, \quad x^{\alpha}\rightarrow x^{\prime\alpha}
= x^{\alpha} + c^{\alpha}t$$

\noindent
with corresponding field transformations
$$v^{\alpha}(t, \vec{x})\rightarrow v^{\prime\alpha}(t, \vec{x}^{\prime})
= v^{\alpha}(t, \vec{x}) - c^{\alpha},$$
$$\hat{v}^{\alpha}(t, \vec{x})\rightarrow \hat{v}^{\prime\alpha}(t,
\vec{x}^{\prime}) = \hat{v}^{\alpha}(t, \vec{x}),$$

\noindent
where $c^{\alpha}$ is a constant velocity. We have then the Lie's derivatives
\ba 
&\delta v^{\alpha}(t, \vec{x}) = 
- c^{\lambda}t\partial_{\lambda}v^{\alpha}(t, \vec{x}) - c^{\alpha},\nonumber \\
&\delta \hat{v}^{\alpha}(t, \vec{x}) = 
- c^{\lambda}t\partial_{\lambda}\hat{v}^{\alpha}(t, \vec{x}).
\label{GLie}
\ea
From (\ref{GLie}) we obtain the  following Ward identity for the functional
generator ${\cal{W}}(J, \hat{J})$
\be
\label{GW}
{\cal{G}}^{\lambda}{\cal{W}}(J, \hat{J}) 
\equiv \int d\hat{x}\left(it\hat{J}^{\alpha}(\hat{x})
{\partial\over \partial x^{\lambda}}{\delta\over \delta\hat{J}^{\alpha}
(\hat{x})} + itJ^{\alpha}(\hat{x})
{\partial\over \partial x^{\lambda}}{\delta\over \delta J^{\alpha}
(\hat{x})} - \hat{J}^{\lambda}(\hat{x})\right){\cal{W}}(J,\hat{J}) = 0. 
\ee
The functional-differential operator ${\cal{G}}^{\lambda}$ carries a
representation of the 
infinitesimal Galilei transformations and it is a generator
of a closed Lie algebra. Indeed the model is also invariant under time and
space translations, given by the Ward identities
\be
\label{HW}
{\cal{H}}{\cal{W}}(J, \hat{J}) 
\equiv \int d\hat{x}\left(i\hat{J}^{\alpha}(\hat{x})
{\partial\over \partial t}{\delta\over \delta\hat{J}^{\alpha}
(\hat{x})} + iJ^{\alpha}(\hat{x})
{\partial\over \partial t}{\delta\over \delta J^{\alpha}
(\hat{x})}\right){\cal{W}}(J, \hat{J}) = 0,
\ee
\be
\label{PW}
{\cal{P}}^{\lambda}{\cal{W}}(J, \hat{J}) 
\equiv \int d\hat{x}\left(i\hat{J}^{\alpha}(\hat{x})
{\partial\over \partial x^{\lambda}}{\delta\over \delta\hat{J}^{\alpha}
(\hat{x})} + iJ^{\alpha}(\hat{x})
{\partial\over \partial x^{\lambda}}{\delta\over \delta J^{\alpha}
(\hat{x})}\right){\cal{W}}(J, \hat{J}) = 0
\ee
which obey the commutation relations
\be 
\label{algebra}
\left[{\cal{G}}^{\lambda}, {\cal{H}}\right] = - {\cal{P}}^{\lambda},
\quad \left[{\cal{G}}^{\lambda}, {\cal{G}}^{\gamma}\right] =
\left[{\cal{G}}^{\lambda}, {\cal{P}}^{\gamma}\right] =
\left[{\cal{P}}^{\lambda}, {\cal{P}}^{\gamma}\right] =
\left[{\cal{P}}^{\lambda}, {\cal{H}}\right] = 0.
\ee
Analogous relations hold for the functionals ${\cal{Z}}(J, \hat{J})$
and $\Gamma(v, \hat{v})$. Introducing the Fourier transform
\be
\tilde{v}^{\alpha}(\hat{q}) = {1\over (2\pi)^2}\int dq^0\int d^3q 
e^{-iq^0t -iq_{\lambda}x^{\lambda}}v^{\alpha}(\hat{x}), ...
\ee
the previous relations are rewritten, for example in terms of the vertex
functional, as
\be
\label{GWF}
-{\cal{G}}^{\lambda}\Gamma(v, \hat{v})
= \int d\hat{q}\left(q^{\lambda}{\partial\over \partial q^0}
\tilde{v}^{\alpha}(\hat{q}){\delta\over \delta\tilde{v}^{\alpha}(\hat{q})} +
q^{\lambda}{\partial\over \partial q^0}
\tilde{\hat{v}}^{\alpha}(\hat{q}){\delta\over
\delta\tilde{\hat{v}}^{\alpha}(\hat{q})}\right)\Gamma(v, \hat{v})
+ {\delta\over \delta \tilde{v}^{\lambda}(\hat{0})}\Gamma(v, \hat{v}) = 0,  
\ee
\be
\label{HWF}
-{\cal{H}}\Gamma(v, \hat{v})
= \int d\hat{q}\left(q^0
\tilde{v}^{\alpha}(\hat{q}){\delta\over \delta\tilde{v}^{\alpha}(\hat{q})} +
q^0\tilde{\hat{v}}^{\alpha}(\hat{q}){\delta\over
\delta\tilde{\hat{v}}^{\alpha}(\hat{q})}\right)\Gamma(v, \hat{v}) = 0,  
\ee
\be
\label{PWF}
-{\cal{P}}^{\lambda}\Gamma(v, \hat{v})
= \int d\hat{q}\left(q^{\lambda}
\tilde{v}^{\alpha}(\hat{q}){\delta\over \delta\tilde{v}^{\alpha}(\hat{q})} +
q^{\lambda}\tilde{\hat{v}}^{\alpha}(\hat{q}){\delta\over
\delta\tilde{\hat{v}}^{\alpha}(\hat{q})}\right)\Gamma(v, \hat{v}) = 0. 
\ee
In order to derive from the previous identities, some general properties of
correlation functions, we introduce the following simplified notation for the
n-points vertex functions. 
\be
\label{nvertice}
\left.
{\delta^{n+m}\Gamma(v, \hat{v})\over \delta\tilde{v}^{\alpha_1}(\hat{q}_1)..
\delta\tilde{v}^{\alpha_n}(\hat{q}_n)\delta\tilde{\hat{v}}^{\alpha_{n+1}}
(\hat{q}_{n+1})..\delta\hat{v}^{\alpha_{n+m}}(0)}
\right|_{v,\hat{v}=0} = \Gamma_{\tilde{v}^{\alpha_1}(\hat{q}_1)..
\tilde{v}^{\alpha_n}(\hat{q}_n)\tilde{\hat{v}}^{\alpha_{n+1}}
(\hat{q}_{n+1})..\hat{v}^{\alpha_{n+m}}(0)}.
\ee
We note that the last field in (\ref{nvertice}) is written in 
configuration space, consequently only the independent momenta appear and no
kinematic $\delta$ singularity is present.

The first non trivial identity is
\be
\label{duetre}
k^{\lambda}{\partial\over \partial k^0}\Gamma_{\tilde{v}^{\alpha}(\hat{k})
\hat{v}^{\beta}(0)} +
\Gamma_{\tilde{v}^{\lambda}(\hat{0})\tilde{v}^{\alpha}(\hat{k})
\hat{v}^{\beta}(0)} = 0, 
\ee
and a consequence of this relation is, for example, 
that the coefficients of the two terms of the convective derivative 
$\partial_t + v^{\alpha}\partial_{\alpha}$ have the same
value. Indeed from power counting and tensorial analysis we set
$$
\Gamma_{\tilde{v}^{\alpha}(\hat{k})\hat{v}^{\beta}(0)}
= \left(-i\sigma_0k^0 + \sigma_{\nu}k^2 +
\Pi(\hat{k})\right)\delta^{\alpha\beta}, $$
and
$$\Gamma_{\tilde{v}^{\lambda}(\hat{p})\tilde{v}^{\alpha}(\hat{k})
\hat{v}^{\beta}(0)} = i\left(\delta^{\beta\lambda}p^{\alpha}
+ \delta^{\beta\alpha}k^{\lambda}\right)\left(\sigma_{\cal{R}}
+ \Sigma(\hat{p}, \hat{k})\right)$$
with $\Pi(\hat{0}) = \Sigma(\hat{0}, \hat{0}) = 0$. From (\ref{duetre}) we
obtain the constraints
\be
\label{convec}
\sigma_0 = \sigma_{\cal{R}} \quad and \quad {\partial\over \partial k^0}
\Pi(\hat{k}) = i\Sigma(\hat{0}, \hat{k}).
\ee
Analogous relations follow from the general
Ward identity of the proper vertex correlation function that can be written as
\be
k^{\lambda}{\partial\over \partial k^0}\Gamma_{\tilde{v}^{\alpha}(\hat{k})
\tilde{v}^{\gamma}(\hat{p})...\hat{v}^{\beta}(0)} +
p^{\lambda}{\partial\over \partial p^0}\Gamma_{\tilde{v}^{\alpha}(\hat{k})
\tilde{v}^{\gamma}(\hat{p})...\hat{v}^{\beta}(0)} + ... +
\Gamma_{\tilde{v}^{\lambda}(\hat{0})\tilde{v}^{\alpha}(\hat{k})
\tilde{v}^{\gamma}(\hat{p})...\hat{v}^{\beta}(0)} = 0. 
\ee

\newsection{The evolution equation}

\noindent
The mechanism of 
energy input in the system, necessary to preserve the stationary
regime, is described by the last term of the Martin, Siggia and Rose action
(\ref{MSRaction}) and it contains a characteristic physical scale. As we
have already observed this scale is given by the parameter $m$ that is the
width, in momentum space, of the two points stirring force correlation function
support and $m = {1\over L}$ where $L$ is of the order of the linear
dimension of the system. In a real physical system an other short-distance
scale, related to the dissipative region, is present; the relevant parameter is
the Kolmogorov's cutoff 
$K_d = {1\over \eta_d}$. The two scales $m$ and $K_d$ are
very important in the physical description of the system. In fact the
dimensionless quantity  
\be
\label{reyn}
\left({K_d\over m}\right)^{4\over 3} = {\cal{R}}
\ee
is the Reynolds number \cite{Com} which is a measure of the nonlinearity
occurring at the scale $L$. If the Reynolds number is very high the system
develops a wide domain 
of the Fourier space ({\it the inertial range}) in which the
non linear terms dominate. 
If we consider the momenta in a region included in this
domain, such that $m\ll q\ll K_d$, both boundary and viscous terms can be
neglected and we expect the 
system to be strongly self similar, homogeneous in
space and isotropic. This 
behavior of the system is interpretable as the presence
of a critical phase in the limit ${\cal{R}}\rightarrow \infty$. In order to
reproduce the previous phenomenology starting from the model described by the
action (\ref{MSRaction}), we consider the Kolmogorov cutoff $K_d$ as a reference
scale and we study the system assuming the other scale $m$ to be very very small
compared with $K_d$. In other words we study the system near the infrared fixed
point (if this fixed point exists) connected to the previous described critical
phase. We expect that, for momenta $q$ such that $m\ll q\ll K_d$, the self
similar behaviour becomes manifest and that correlation functions obey, with
good approximation, the behaviour given by dimensional analysis. In particular
the energy spectrum $E(q)$ must follow a power law of the form \cite{KO}
\be
\label{Kspet}
E(q) = C_K{\cal{E}}^{2\over 3}q^{-{5\over 3}},
\ee
where $C_K$ is a constant of the order of unity and the
renormalization group methods allow a quantitative estimation of
the $C_K$ Kolmogorov constant. 
Experimental tests prove that the scaling behavior of correlation
functions of order higher then 3 has some corrections. Such corrections are
generally linked to the intermittent behavior of turbulent systems \cite{UF}, i.
e. to strong, nonlinear, rare events that are usually associated with the
instantonic configurations. This aspect is not discussed in the approach here
analysed.

In order to obtain an evolution equation for the model (\ref{MSRaction}) we
introduce the six components field
$\tilde{\Psi}^a(\hat{q})\equiv\left(\tilde{v}^{\alpha}(\hat{q}),
\tilde{\hat{v}}^{\alpha}(\hat{q})\right)$ and the density of lagrangian
\be
\label{MSRlagr}
{\cal{L}}(q; m) = {\cal{L}}^0(q) + {1\over 2}\tilde{\Psi}^a(-\hat{q})
R_{ab}(q; m)\tilde{\Psi}^b(\hat{q})
\ee
\noindent
where ${\cal{L}}^0(q)$ is the original Navier-Stokes lagrangian (independent
of the scale $m$) without stirring force, which is totally included in
the last term where the matrix $R_{ab}(q; m)$ is given by
\be
\label{RF}
R_{ab}(q; m) = \left(\begin{array}{cc}
0&  0\\  
0&  2ih(q; m)P^{\alpha\beta}(q)\end{array}\right). 
\ee
Concerning the function $h(q; m)$, it must satisfy the physical constraint 
(\ref{E}) and its momentum spectrum should be of width $m$.
A possible choice is
\be
\label{HN}
h(q; m) = {D_0\over m^3}\chi\left({q\over m}\right),
\ee
where, for example
$$\chi\left({q\over m}\right) = \xi\left({q\over m}\right)e^{-{q^2\over m^2}}.$$
$D_0$ is a dimensional parameter fixed by the constraint (\ref{E}), once the
homogeneous function $\chi\left({q\over m}\right)$ is specified. The function
(\ref{HN}) is very similar to that chosen by P. Brax \cite{PB} and as the
parameter $m$ is different from zero, the density of lagrangian
(\ref{MSRlagr}) corresponds to a model with a Galilei-invariant
infrared regularization. The ultraviolet
convergence is assured  by the same function (\ref{HN}) and by the existence
of a physical cutoff $\Lambda_0 > K_d$ where, for momenta $q\sim\Lambda_0$,
all statistical fluctuations rapidly vanish due to the dissipative effects.

Taking into account that all the $m$ dependence of the functional 
$\Gamma(\Psi; m)$ ($\Gamma(\Psi; m)\equiv\Gamma(v,\hat{v})$) 
is contained in the function $h(q; m)$,  one easily \cite{J.Pol}, \cite{BDM},
\cite{JW} derives an evolution equation in $m$
$$m\partial_m\left(\Gamma(\Psi; m) - {1\over 2}\int
d\hat{q}\tilde{\Psi}^a(-\hat{q})R_{ab}(q; m)\tilde{\Psi}^b(\hat{q})\right)$$
\be
\label{evolut}
= -{i\over 2}\int {d\hat{q}\over (2\pi)^4}m\partial_mR_{ab}(q; m)
\Delta_{\Psi_b\Psi_c}(\hat{q}; m){\delta^2\bar{\Gamma}(\Psi; m)\over
\delta\tilde{\Psi}_c(\hat{q})\delta\tilde{\Psi}_l(-\hat{q})}
\Delta_{\Psi_l\Psi_a}(\hat{q}; m),  
\ee
where $\Delta_{\Psi_b\Psi_c}(\hat{q}; m)$ is the full propagator which, in term
of the two points proper correlation functions, is given by the matrix
\be
\label{propag}
\Delta_{\Psi_a\Psi_b}(\hat{q}; m) =
\Gamma^{-1}_{\tilde{\Psi}_a(\hat{q})\Psi_b(0)}
= \left(\begin{array}{cc}
-{\Gamma_{\tilde{\hat{v}}^{\gamma}(\hat{q})\hat{v}^{\rho}(0)}P^{\rho\sigma}(q)
\over \Gamma_{\tilde{v}^{\alpha}(\hat{q})\hat{v}^{\gamma}(0)}
\Gamma_{\tilde{\hat{v}}^{\sigma}(\hat{q})v^{\beta}(0)}}&  
{P^{\alpha\sigma}(q)\over
\Gamma_{\tilde{v}^{\sigma}(\hat{q})\hat{v}^{\beta}(0)}}\\   
{P^{\alpha\sigma}(q)\over
\Gamma_{\tilde{\hat{v}}^{\sigma}(\hat{q})v^{\beta}(0)}}&  0\end{array}\right).
\ee
The auxiliary functional ${\delta^2\bar{\Gamma}(\Psi; m)\over
\delta\tilde{\Psi}_a(\hat{q})\delta\tilde{\Psi}_b(\hat{q}^{\prime})}
\equiv \bar{\Gamma}_{ab}(\hat{q}, \hat{q}^{\prime}; m)$ is defined by the
integral equation
\be
\bar{\Gamma}_{ab}(\hat{q}, \hat{q}^{\prime}; m) =  
{\Gamma}^{int}_{ab}(\hat{q}, \hat{q}^{\prime}; m)
-\int d\hat{q}^{\prime\prime}\bar{\Gamma}_{cb}(-\hat{q}^{\prime\prime}, 
\hat{q}^{\prime}; m)\Delta_{\Psi_d\Psi_c}(\hat{q}^{\prime\prime}; m)
\Gamma^{int}_{ad}(\hat{q}, \hat{q}^{\prime\prime}; m)
\ee
where $\Gamma^{int}_{ab}(\hat{q}, \hat{q}^{\prime}; m)$ is obtained by the
functional
\be
\Gamma^{int}\left(\Psi; m\right) = \Gamma\left(\Psi; m\right)
- {1\over 2}\int d\hat{q}\tilde{\Psi}_a(-\hat{q})
\Gamma_{\tilde{\Psi}_a(\hat{q})\Psi_b(0)}\tilde{\Psi}_b(-\hat{q}),
\ee
i, e. the generator $\Gamma\left(\Psi; m\right)$ without the
bilinear terms. In eq.(\ref{propag}) we have assumed  that 
$\Gamma_{v^{\alpha}v^{\beta}} = 0$; in perturbation theory all vertex functions
containing only $v^{\alpha}$ field vanish, but this property is an exact
consequence of (\ref{evolut}) and the boundary conditions (see appendix A).
From (\ref{evolut}) we obtain the equation for the vertex correlation
functions. In order to tackle specific problems, consider the flow equations
for the two points vertex functions $\Gamma_{\tilde{v}^{\alpha}(\hat{p})
\hat{v}^{\beta}(0)}$ and $\Gamma_{\tilde{\hat{v}}^{\alpha}(\hat{p})
\hat{v}^{\beta}(0)}$; they are

$$m\partial_m\Gamma_{\tilde{v}^{\alpha}(\hat{p})\hat{v}^{\beta}(0)}
= \int {d\hat{q}\over (2\pi)^4}m\partial_mh(q; m)
\left(\Delta_{\hat{v}^{\gamma}v^{\rho}}(\hat{q}; m)
\Gamma_{\tilde{v}^{\rho}(\hat{q})\tilde{v}^{\delta}(-\hat{q})
\tilde{v}^{\alpha}(\hat{p})\hat{v}^{\beta}(0)}
\Delta_{v^{\delta}\hat{v}^{\gamma}}(\hat{q}; m)\right.$$
$$- \Delta_{\hat{v}^{\gamma}v^{\rho}}(\hat{q}; m)
\Gamma_{\tilde{v}^{\rho}(\hat{q})
\tilde{v}^{\alpha}(\hat{p})\Psi^s(0)}
\Delta_{\Psi^s\Psi^r}(\hat{q} + \hat{p}; m)
\Gamma_{\tilde{\Psi}^r(\hat{q}+\hat{p})
\tilde{v}^{\delta}(-\hat{q})\hat{v}^{\beta}(0)}
\Delta_{v^{\delta}\hat{v}^{\gamma}}(\hat{q}; m)$$
\be
\label{vhv}
\left.- \Delta_{\hat{v}^{\gamma}v^{\rho}}(\hat{q}; m)
\Gamma_{\tilde{v}^{\rho}(\hat{q})
\tilde{\Psi}^s(\hat{p}-\hat{q})\hat{v}^{\beta}(0)}
\Delta_{\Psi^s\Psi^r}(\hat{p} - \hat{q}; m)
\Gamma_{\tilde{\Psi}^r(\hat{q}-\hat{p})
\tilde{v}^{\alpha}(\hat{p})v^{\delta}(0)}
\Delta_{v^{\delta}\hat{v}^{\gamma}}(\hat{q}; m)\right)
\ee

\noindent
and
$$m\partial_m\left(\Gamma_{\tilde{\hat{v}}^{\alpha}(\hat{p})\hat{v}^{\beta}(0)}
-2ih(p; m)P^{\alpha\beta}(p)\right)$$
$$= \int {d\hat{q}\over (2\pi)^4}m\partial_mh(q; m)
\left(\Delta_{\hat{v}^{\gamma}v^{\rho}}(\hat{q}; m)
\Gamma_{\tilde{v}^{\rho}(\hat{q})\tilde{v}^{\delta}(-\hat{q})
\tilde{\hat{v}}^{\alpha}(\hat{p})\hat{v}^{\beta}(0)}
\Delta_{v^{\delta}\hat{v}^{\gamma}}(\hat{q}; m)\right.$$
\be
\label{hvhv}
\left.- \Delta_{\hat{v}^{\gamma}v^{\rho}}(\hat{q}; m)
\Gamma_{\tilde{v}^{\rho}(\hat{q})
\tilde{\hat{v}}^{\alpha}(\hat{p})\Psi^s(0)}
\Delta_{\Psi^s\Psi^r}(\hat{q} + \hat{p}; m)
\Gamma_{\tilde{\Psi}^r(\hat{q}+\hat{p})
\tilde{v}^{\delta}(-\hat{q})\hat{v}^{\beta}(0)}
\Delta_{v^{\delta}\hat{v}^{\gamma}}(\hat{q}; m)\right).
\ee

\noindent
In the r. h. s. of the previous equations there appear three and four points
vertex functions that are defined by flow equations involving higher
correlations thus producing an infinite
system. We
will analyze the fundamental point of truncating this system in another section,
studying here only the general properties and the boundary conditions. 

If there is an infrared fixed
point in the limit $m\rightarrow 0$, it is convenient to study the system near
this point. For this reason we measure all momenta in terms of the
spectral parameter $m$ and set
\be
p^0=\tau m^2,\quad  \left|\vec{p}\right| = p = xm, \quad q^0 = \eta m^2,
\quad \left|\vec{q}\right| = q = ym, ... 
\ee
We suppose that the spectral parameter $m$ is very small compared with the
Kolmogorov cutoff $K_d$. The 
space of the $x$-like variables is then divided into
three regions. In the first region (the infrared region) the p momenta are 
smaller then $m$ and 
$$x<1.$$
Here all correlation functions are strongly dependent on the form of the
stirring force. 
In the second region (the inertial region) the $p$ momenta are
finite and $m$ is very small compared with these momenta. Here we have
$$x\gg 1.$$

\noindent
In this region we expect a strong self-similarity behaviour of the system,
provided that
$$x\ll {K_d\over m}.$$
The last region is the dissipative region, where
$$x>{K_d\over m}$$
The boundary conditions for eq.s(\ref{vhv}) and (\ref{hvhv}), and for all
possible flow equation, are obtained in this region by the consideration 
that for  
$$x = x_0 = {\Lambda_0\over m}\sim\infty, \quad where \quad
\Lambda_0>K_d$$

\noindent
all statistical fluctuations vanish as a consequence of the dissipative effects.
In other words for any external variable $x$ of the order ${\Lambda_0\over m}
\sim\infty$ all correlation vertex functions with a field content different from
the vertices of the original Navier-Stokes action, without the stirring force
term, vanish. Therefore the boundary value of the effective action is the
classical Navier-Stokes action.

From equations (\ref{vhv}), (\ref{hvhv}) and the previous considerations it
is possible to extract some general informations about the vertex correlation
functions. Let us consider the two point vertex functions
$\Gamma_{\tilde{v}^{\alpha}(\hat{p})\hat{v}^{\beta}(0)}$, that we rewrite, in
terms of the rescaled variables and taking into account the power counting
analysis, as
\be
\label{4vhv} 
\Gamma_{\tilde{v}^{\alpha}(\hat{p})\hat{v}^{\beta}(0)}
= \Gamma^m_{\tilde{v}^{\alpha}(\tau, x)\hat{v}^{\beta}(0)}
=m^2\left[-il\left(\tau, {D_0^{1\over 2}\over m^2}\right)\tau + f\left(\tau,
x, {D_0^{1\over 2}\over m^2}\right)x^2\right]P^{\alpha\beta}(p),
\ee
\be
\label{4hvhv}
\Gamma_{\tilde{\hat{v}}^{\alpha}(\hat{p})\hat{v}^{\beta}(0)} =
\Gamma^m_{\tilde{\hat{v}}^{\alpha}(\tau, x)\hat{v}^{\beta}(0)} = 2i{D_0\over
m^3}\left[\chi(x) + M\left(\tau, x, {D_0^{1\over 2}\over m^2}\right)
\right]P^{\alpha\beta}(p).
\ee

\noindent
These functions are defined, at $m\not= 0$, for all values of the variable 
$\tau$ and $x$. In particular they satisfy
$\Gamma^m_{\tilde{v}^{\alpha}(0, 0)\hat{v}^{\beta}(0)} = 0$ and
$M(\tau, 0,..) = 0$ for any $m\not= 0$.
It is possible also to see that all vertex functions (with the
exclusion of $\Gamma_{\hat{v}\hat{v}}$) are almost proportional to the momenta
of the $\hat{v}$ fields. If we go back to equation (\ref{hvhv}) we see that
some inconsistencies are present: indeed the r. h. s. of this equation, for 
$x=0$ is different from zero due to infrared divergences of the full
propagator. This apparent inconsistency is caused by the fact that in the
function $h(x; m) 
={D_0\over m^3}\chi(x)$ the value $x = 0$ is not accessible due
to the finite size of the system and therefore we must define
\be
\label{origine}
\chi(0) = 0.
\ee
\noindent
The equations (\ref{hvhv}), (\ref{vhv}) are defined for arbitrary values of the
variable $x$ and the previous condition for the function $h(x; m)$ at $x=0$ is a
physical regularization connected with the infinite volume limit. Taking into
account condition (\ref{origine}) we obtain a flow equation for the quantity
$\Gamma^m_{\tilde{v}^{\alpha}(\tau, 0)\hat{v}^{\beta}(0)} = -im^2l(\tau,..)\tau$
$$\left(m\partial_m -
2\tau\partial_{\tau}\right)\left[m^2l\left(\tau,
{D_0^{1\over 2}\over m^2}\right)\tau\right] = 0
\rightarrow\left(m\partial_m - 2\tau\partial_{\tau}\right)l\left(\tau,
{D_0^{1\over 2}\over m^2}\right) = 0$$

\noindent
which has the general integral

$$l = l\left(D_0^{-{1\over 2}}m^2\tau\right).$$

\noindent
But the asymptotic boundary conditions in $x = {\Lambda_0\over m}$
(and $\tau = {\Lambda^2_0\over m^2}$) enforce
$l\left(D_0^{-{1\over 2}}m^2\tau\right) = 1$, independently of the values of the
$m$ variable. Therefore we have the exact result
$$l(\tau, ..)\equiv 1.$$
Thus the two points vertex function is rewritten as
\be
\label{2punti}
\Gamma^m_{\tilde{v}^{\alpha}(\tau, x)\hat{v}^{\beta}(0)}
=m^2\left[-i\tau + f\left(\tau,
x, {D_0^{1\over 2}\over m^2}\right)x^2\right]P^{\alpha\beta}(p).
\ee
From the Ward identity (\ref{duetre}) and from (\ref{convec}) we have another
exact result concerning the three points vertex function 
$\Gamma_{\tilde{v}^{\lambda}(\hat{q})\tilde{v}^{\alpha}(\hat{p}-\hat{q})
\hat{v}^{\beta}(0)}$ which is
$$
\Gamma_{\tilde{v}^{\lambda}(\hat{q})\tilde{v}^{\alpha}(\hat{p}-\hat{q})
\hat{v}^{\beta}(0)} =
\Gamma^m_{\tilde{v}^{\lambda}(\eta, y)\tilde{v}^{\alpha}(\tau-\eta, x-y)
\hat{v}^{\beta}(0)}$$
\be
\label{3punti} 
= imx\left(\delta^{\beta\lambda}n^{\alpha} +
\delta^{\beta\alpha}n^{\lambda}\right)\left(1 + \Sigma(\tau, \eta, x, y,
..)\right) 
\ee
where $p^{\alpha} = mxn^{\alpha}$. These exact results, which are in agreement
with the non renormalizability theorems of this model, were already known in
perturbation theory.

\newsection{The approximation scheme}

\noindent
In order to study the structure functions of the two points correlations
functions, let us consider eq.s (\ref{vhv}), (\ref{hvhv}). As we have already
observed they are equivalent to an infinite system of flow equations; 
therefore a truncation procedure is needed. The
important observation is that the correlation vertex functions, which have a
different field content with respect to the original classical Navier-Stokes
action, are relevant in the infrared region, i,e. when the x-like variables are
near unity, but they are less relevant in the
inertial region where $x\gg 1$. 
The physical reason for this expected behavior is
due to the statistical fluctuations induced  by the non local vertex
$2ih(q; m)P^{\alpha\beta}(q)\hat{v}^{\alpha}(-\hat{q})\hat{v}^{\beta}(\hat{q})$
which has a very concentrated Fourier spectrum in the long wave length region.
The non linear terms transfer, by the cascade mechanism, these fluctuations to
all  regions of the Fourier spectrum, but the speed of this transfer depends 
on the coupling of the non linear term with the degrees of freedom of the
system; in other words on the Reynolds number. If the Reynolds number is very
large the statistical fluctuations are swiftly transferred to a spectral
region where the dissipative effects are dominant and,
in order to preserve the stationary regime, the corresponding amplitudes are
decreasing for increasing momenta. A more specific analysis follows from the
connection between the parameters $D_0$, $m$, the quantities related to the
regime of the physical system such as the Reynolds number, and the rate of power
dissipated by a mass unit $\cal{E}$ of fluid. From the normalization condition
(\ref{E}) we have for the dimensionless quantity $D_0\over m^4$
$${D_0\over m^4} = {\pi^2\over \int dy y^2\chi(y)}{\langle{\cal{E}}\rangle
\over m^4}.$$
The Kolmogorov's cutoff is defined by $K_d^4 = {{\cal{E}}\over \nu^3}$ with
$\nu$ the kinematic viscosity. Recalling then (\ref{reyn}) the following
relation is obtained
\be
\label{DR}
{D_0\over m^4} = {\pi^2\over \int dy y^2\chi(y)}\nu^3{\cal{R}}^3.
\ee
As we see the statistical fluctuations vanish for ${\cal{R}}\rightarrow 0$
according to the fact that, in this case, no transfer of energy in a spectral
region different from m is possible and the energy pumped into the system at
this scale is immediately dissipated (indeed for ${\cal{R}}\rightarrow 0$
$m\gg K_d$).

If we analyze in more detail eq.s (\ref{vhv}), (\ref{hvhv}) we find in the 
r.h.s. the three and four points proper vertex correlation functions  
\ba
&\Gamma_{\tilde{v}^{\rho}(\hat{q})\tilde{v}^{\alpha}(\hat{p})
\hat{v}^s(0)}, \quad
\Gamma_{\tilde{v}^r(\hat{q}+\hat{p})\tilde{v}^{\delta}(-\hat{q})
\hat{v}^{\beta}(0)},
\quad \Gamma_{\tilde{v}^{\rho}(\hat{q})\tilde{\hat{v}}^{\alpha}(\hat{p})
\hat{v}^s(0)}\\
\label{3V}
&\Gamma_{\tilde{v}^{\rho}(\hat{q})\tilde{v}^{\delta}(-\hat{q})
\tilde{v}^{\alpha}(\hat{p})\hat{v}^{\beta}(0)}, \quad
\Gamma_{\tilde{v}^{\rho}(\hat{q})\tilde{v}^{\delta}(-\hat{q})
\tilde{\hat{v}}^{\alpha}(\hat{p})\hat{v}^{\beta}(0)}
\label{4V}
\ea
and the two points proper correlation functions appear in the full
propagators as results from (\ref{propag}). Concerning these last correlation
functions we have a first observation about the function
$\Gamma_{v^{\alpha}\hat{v}^{\beta}}$. From 
(\ref{2punti}) we rewrite the function
$f(\tau, x, ...)$ isolating the terms which depend on the $\tau$ variable i,e.

$$f\left(\tau, x; {D_0\over m^4}\right) = f\left(x; {D_0\over m^4}\right)
+ f_{\tau}\left(\tau, x; {D_0\over m^4}\right).$$

\noindent
From the Ward identity (\ref{convec}) and (\ref{3punti}) we obtain
\be
\label{t3}
{\partial\over \partial\tau}f_{\tau}\left(\tau, x; {D_0\over m^4}\right)x^2
=i\Sigma\left(\tau,0,x,0; {D_0\over m^4}\right)
\ee
i. e. $f_{\tau}\left(\tau, x; {D_0\over m^4}\right)x^2$ has the same scaling
behavior of the statistical corrections to the three points classical vertex.
Therefore, for the previous argument, we expect that $f_{\tau}$ decreases faster
then $f$. In a completely similar way we see that all $\tau$ dependent
corrections of the vertex  functions with $n$ $v$ and $m$ $\hat{v}$ fields, are
connected, by the Ward identities, to the functions with $n+1$ $v$ and $m$
$\hat{v}$ fields. Writing
\be
\label{gerarchia}
f_{\tau}\left(\tau, x; {D_0\over m^4}\right)x^2 
= -i\tau g\left(x; {D_0\over m^4}\right) +
\tau^2 g_{(2)}\left(x; {D_0\over m^4}\right) + ... +
\tau^n g_{(n)}\left(x; {D_0\over m^4}\right) + ...,
\ee
from the last observations we have that $g\left(x; {D_0\over m^4}\right)$
is related to three point vertex statistical corrections, 
$g_{(2)}\left(x; {D_0\over m^4}\right)$ is related to the four point vertex
and so on. Taking into account
the previous considerations we can write  the flow equation for  $f(x; {D_0\over
m^4})$, where we analyze only the leading contributions appearing in the triple
vertex. In other words considering the vertex functions that
appear in the r. h. s.
of equation (\ref{vhv}), i. e. the three and four points function
$\Gamma^m_{\tilde{v}^{\lambda}(\eta, y)\tilde{v}^{\alpha}(-\eta, x-y)
\hat{v}^{\beta}(0)}$,  
$\Gamma^m_{\tilde{v}^{\rho}(\eta,
y)\tilde{v}^{\delta}(-\eta,-y)\tilde{v}^{\alpha}(0,x)\hat{v}^{\beta}(0)}$, we
separate the leading contribution from the statistical correction by writing the
vertex $\Gamma^m_{vv\hat{v}}$ as
\be
\label{leading}
\Gamma^m_{vv\hat{v}}=\Gamma^{(0)}_{vv\hat{v}} +
\left(\Delta\Gamma\right)_{vv\hat{v}}
\ee
and we suppose, for $x\gg 1$, the following approximations
\be
\label{appr1}
\displaystyle\lim_{x\gg 1}
{\Gamma^m_{vv\hat{v}}}\sim \Gamma^{(0)}_{vv\hat{v}}\quad and
\quad \displaystyle\lim_{x\gg 1}
{\Gamma^m_{vvv\hat{v}}}\sim 0.
\ee
The same analysis is performed for eq.(\ref{hvhv}) where the vertices
involved are 
$\Gamma^m_{\tilde{v}^{\lambda}(\eta,y)\tilde{v}^{\rho}(-\eta,x-y)
\hat{v}^{\beta}(0)}$, $\Gamma^m_{\tilde{v}^{\lambda}(\eta,y)
\tilde{\hat{v}}^{\rho}(-\eta,x-y)\hat{v}^{\beta}(0)}$, 
$\Gamma^m_{\tilde{v}^{\lambda}(\eta,y)\tilde{v}^{\rho}(-\eta,-y)
\tilde{\hat{v}}^{\alpha}(0,x)\hat{v}^{\beta}(0)}$, with the 
approximations
\be
\label{appr2}
\displaystyle\lim_{x\gg 1}
{\Gamma^m_{v\hat{v}\hat{v}}}\sim 0\quad and
\quad 
\displaystyle\lim_{x\gg 1}
{\Gamma^m_{vv\hat{v}\hat{v}}}\sim 0.
\ee
In this way
eq.s (\ref{vhv}), (\ref{hvhv}) give a closed system of differential equations. 
The approximation here proposed can also be obtained by a suitable field
truncation, preserving the 
Galilei-invariance, in the effective irreducible action
\cite{PT}. 

In order to write, from (\ref{vhv}) and (\ref{hvhv}),  the equations for the
structure functions $f\left(x; {D_0\over m^4}\right)$ and 
$M\left(x; {D_0\over m^4}\right)$ we define
\be
\label{fstr}
f\left(x; {D_0\over m^4}\right) \equiv \left(D_0\over m^4\right)^{1\over
3}\phi_m(x), 
\ee
\be
\label{mstr}
M\left(x; {D_0\over m^4}\right) \equiv M_m(x). 
\ee
and after straightforward calculations we obtain
\ba
&\left(m\partial_m - x\partial_x - {4\over 3}\right)\phi_m(x) \nonumber\\
&= 
\int {y^2dy\over 4\pi^2}\int_{-1}^1dt{(1-t^2)\left(t\left({y\over x}-2{x\over
y}\right)-{x^2\over y^2}\right)(3 + y\partial_y)\chi(y)\over
\phi_m(y)\left(x^2+y^2+2txy\right)\left[\phi_m\left(\sqrt{x^2+y^2+2txy}\right)
\left(x^2+y^2+2txy\right) + \phi_m(y)y^2\right]},
\label{phieq} 
\ea
\ba
&\left(m\partial_m - x\partial_x - 3\right)M_m(x) \nonumber \\
&=
\int {y^2dy\over 4\pi^2}\int_{-1}^1dt{(1-t^2)x^2\left(1+2t^2 +{x^2\over y^2}
+3t{x\over y}\right)(3 + y\partial_y)\chi(y)
\left(\chi_m\left(\sqrt{x^2+y^2+2txy}\right) 
+ M_m\left(\sqrt{x^2+y^2+2txy}\right)\right)\over \phi_m(y)
\phi_m\left(\sqrt{x^2+y^2+2txy}\right)\left(x^2+y^2+2txy\right)^2
\left[\phi_m\left(\sqrt{x^2+y^2+2txy}\right)
\left(x^2+y^2+2txy\right) + \phi_m(y)y^2\right]}.\nonumber\\
\label{Meq} 
\ea
Concerning the boundary and initial conditions for (\ref{phieq}) and (\ref{Meq})
we must consider that these equations are valid only in the inertial
region, i.e. for $x\gg 1$, provided $x\ll {K_d\over m}\equiv x_d$. As the
parameter $m$ is very small compared with $K_d$ (we are studying the system near
the infinite volume limit) it follows that $x_d\sim \infty$ is a good
approximation. Recalling that the Kolmogorov scale $K_d$ is related to the
viscosity and to the rate of energy dissipation by the relation $K_d\approx
\left({{\cal{E}}\over \nu^3}\right)^{1\over 4}$ \cite{Com}, the previous
approximation, for $m\not= 0$, is equivalent to $\nu\sim 0$. In conclusion we
extrapolate the following boundary conditions:
\be
\label{bound}
\phi_m(\infty) = M_m(\infty) = 0, \quad for\quad m\not= 0.
\ee
It is clear that (\ref{phieq}) and (\ref{Meq}) with (\ref{bound}) enable us
to describe the physical system in the dissipative region. Considering that we
are studying the 
system near the supposed fixed point in the infinite volume limit,
the initial conditions are given for $m\rightarrow 0$. As we have already 
observed our model corresponds to a regularized theory and, for vanishing
regularization parameter $m$, the infrared divergences, due to the infinite
volume limit, will appear 
again. The choices (\ref{fstr}), (\ref{mstr}) fix these
divergences according to the power counting and dimensional analysis; in
$\phi_m(x)$ and $M_m(x)$ no explicit dependence on the parameter $m$ is present.
It is 
also evident that for $x\sim 1$ the considered equations are inadequate for
the description of the physical system.

The meaning of our approximation scheme is evident by observing 
eq.s(\ref{phieq}), (\ref{Meq}) which correspond to the first step of the loop
expansion. However these equations contain a nonperturbative contribution
concerning the structure functions (\ref{fstr}) and (\ref{mstr}) which are
solutions of the 
same equations with the initial conditions for $m\rightarrow 0$.

The next step requires the one loop calculation of the vertex functions
$\left(\Delta\Gamma\right)_{vv\hat{v}}$, $\Gamma^m_{vvv\hat{v}}$,
$\Gamma^m_{v\hat{v}\hat{v}}$ and $\Gamma^m_{vv\hat{v}\hat{v}}$ still using the
ERG containing in the r. h. s. only the leading terms and the structure functions
(\ref{fstr}) and (\ref{mstr}). 
The two loops contributions follow inserting these
vertices in 
the original equations (\ref{vhv}), (\ref{hvhv}). This procedure is
extendible to 
an arbitrary order and it contains a non perturbative input given by
the structure functions (\ref{fstr}) and (\ref{mstr}). Naturally this scheme is
meaningful only if the two loop are smaller then the one loop contributions, and
so on. We 
also observe that, in our approach, no particular hypothesis concerning
the effective action is needed and this result follows from a self-consistency 
argument of our model.

\newsection{Numerical results}

\noindent
In this section we report some numerical results concerning the solutions of the
system (\ref{phieq}), (\ref{Meq}). The analysis takes into account different
choices of the stirring force, all satisfying the spectral condition discussed in
the previous sections. By 
the expected universality in the inertial range we must
verify that the exact shape of the noise correlator is not very
important. We consider a $\chi(x)$ function with the general form
\be
\label{noispec}
\chi(x) = x^se^{-x^{2n}}
\ee
where $s$ and $n$ are positive integers. Eq.s (\ref{phieq}), (\ref{Meq}) are
solved with an explicit finite-difference scheme starting at the boundary
conditions (\ref{bound}) and by the initial conditions which require that no
explicit m-dependence is present for $m\rightarrow 0$ (i. e. for the system near
the critical point). The numerical computations show that the solutions maintain
explicit independence on the scale-parameter $m$. The function $\phi(x)$ and
$M(x)$ are computed 
in the range $10^{-1}<x<10^2$. The inertial range corresponds
to $x\gg 1$ and in this region the function $\phi(x)$ and $M(x)$ reach a scale
invariant regime. Indeed, referring to the parametrization  
\be
\phi(x) = \sigma x^{-{4\over 3}+\alpha} \quad and \quad M(x) = \gamma x^{-3
+\beta}, 
\ee
\begin{figure}
\begin{center}
\includegraphics[angle=270, width=10cm]{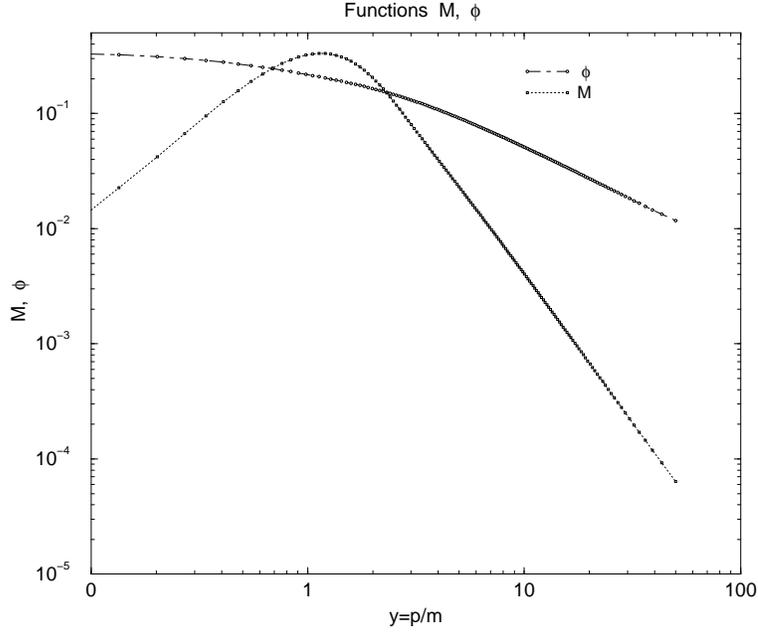}
\end{center}
\caption{$\phi$ and $M$ functions} 
\label{strut}
\end{figure}
the numerical computation  gives 
for $\alpha$ and $\beta$ a value $\sim +0.37$ in
the region $x>1$ and the 
function $\phi(x)$ and $M(x)$ are graphically represented
in  Fig.(\ref{strut}). The detailed numerical values of the parameters related
to the functions $\phi(x)$ and $M(x)$ obtained with different noise functions
$\chi$ are summarized in table (1) and the substantial independence of the
exponents $-{4\over 3} +\alpha$ and $-3 +\beta$ on the particular form of
$\chi(x)$ appears very clearly.

\begin{center}
\begin{tabular}{|l|c|c|c|c|c|}
\hline
$\chi(x)$ & 
$\sigma$ ($\pm 0.001$) & $-{4\over 3} +\alpha (\pm 0.0005)$ & $\gamma$
($\pm 0.001$) & $-3 +\beta$ ($\pm 0.0005$) & ${D_0\over
m^4}\left(\nu{\cal{R}}\right)^{-3}$ 
\\ \hline\hline
$x^2e^{-x^2}$ & 0.899 & -0.924 & 20.00 & -2.590 & 14.85 \\
$x^2e^{-x^4}$ & 0.900 & -0.935 & 22.49 & -2.601 & 43.55 \\
$x^2e^{-x^6}$ & 0.935 & -0.930 & 26.02 & -2.630 & 52.46 \\
$x^4e^{-x^2}$ & 0.924 & -0.933 & 29.63 & -2.599 & 5.94 \\
$x^4e^{-x^4}$ & 0.994 & -0.929 & 28.97 & -2.595 & 42.95 \\
$x^6e^{-x^6}$ & 1.02 & -0.930 & 33.55 & -2.596 & 66.81 \\
$x^8e^{-x^8}$ & 1.03 & -0.930 & 35.93 & -2.633 & 88.82 \\
$x^{10}e^{-x^{10}}$ & 1.08 & -0.929 & 38.05 & -2.595 & 109.97 \\
$x^{12}e^{-x^{12}}$ & 1.09 & -0.931 & 38.41 & -2.597 & 109.97 \\
$e^{-{1\over x^2}}e^{-x^2}$ & 0.942 & -0.922 & 27.70 & -2.588 &54.86 \\
\hline
\end{tabular}
\end{center}
\centerline{Table 1: Parameter of the functions $\phi(x)$ and $M(x)$ for various
stirring forces.}

\noindent
A significative check of our analysis is given by the computation of the  energy
spectrum. From 
$$
E(\hat{x}; m) = -{1\over 2}\left.{\delta^2{\cal{Z}}\over 
\delta\hat{J}^{\alpha}(\hat{x})\delta\hat{J}_{\alpha}(\hat{x})}
\right|_{{\cal{J}}=0} = {1\over 2}\langle u^{\alpha}(t, \vec{x})
u_{\alpha}(t, \vec{x})\rangle_c
$$
\be
= {1\over 2(2\pi)^4}\int dq^0q^2dqd\Omega_q{1\over 
\Gamma^m_{\tilde{u}^{\alpha}\hat{u}^{\gamma}}(\hat{q})}
\Gamma^m_{\tilde{\hat{u}}^{\gamma}\hat{u}^{\rho}}(\hat{q})
{1\over 
\Gamma^m_{\tilde{\hat{u}}^{\rho}u^{\alpha}}(\hat{q})} 
= \int dqE(q; m)
\ee
we obtain
\be
\label{Espet}
E(q; m) = {1\over 2\pi^2}q^2{M({q\over m})\over \phi({q\over m})q^2}m^{-{5\over
3}}D_0^{2\over 3} = {1\over 2}\left({\gamma\over \sigma}\right)\left({1\over 
\int dy y^2\chi(y)}\right)^{2\over 3}{\cal{E}}^{2\over 3}q^{-{5\over 3}}\left(
{q\over m}\right)^{\beta - \alpha}. 
\ee 
Recalling (\ref{Kspet}) the Kolmogorov constant $C_K$ is given by the
$q$-independent  part of (\ref{Espet}) i.e.
\be
\label{CK}
C_K = {1\over 2}\left({\gamma\over \sigma}\right)\left({1\over 
\int dy y^2\chi(y)}\right)^{2\over 3}.
\ee
The energy power behavior in terms of the q-variable is given by the exponent 
$\xi = -{5\over 3}+\beta -\alpha$. 
The numerical values of $C_K$ and $\xi$, computed with different noise functions
$\chi(x)$ are summarized in the following table
\begin{figure}
\begin{center}
\includegraphics[angle=270, width=10cm]{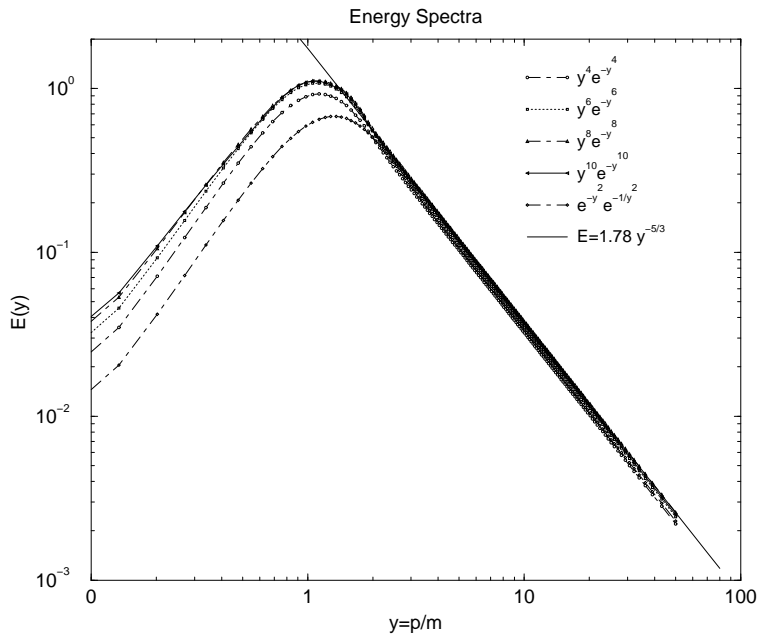}
\end{center}
\caption{Energy spectra for various $\chi(x)$ functions} 
\label{spettr}
\end{figure}

\begin{center}
\begin{tabular}{|l|c|c||l|c|c|}
\hline
$\chi(x)$ & $\xi$ ($\pm 0.001$) & $C_K (\pm 0.002)$ & $\chi(x)$ &
$\xi$ ($\pm 0.001$) & $C_K (\pm 0.002)$ \\
\hline\hline
$x^2e^{-x^2}$ & -1.666 & 1.124 & $x^4e^{-x^4}$ & -1.666 & 1.461 \\
$x^2e^{-x^4}$ & -1.667 & 1.267 & $x^6e^{-x^6}$ & -1.666 & 1.660 \\
$x^2e^{-x^6}$ & -1.666 & 1.417 & $x^8e^{-x^8}$ & -1.666 & 1.767 \\
$e^{-{1\over x^2}}e^{-x^2}$ & -1.667 & 1.489 & $x^{10}e^{-x^{10}}$ & -1.666 &
1.784 \\
$x^4e^{-x^2}$ & -1.667 & 1.624 & $x^{12}e^{-x^{12}}$ & -1.666 & 1.785 \\
\hline
\end{tabular}
\end{center}

\centerline{Table 2: Slope $\xi$ and Kolmogorov constant $C_K$ for various
stirring forces}

\noindent
Looking at table (2) the exponent $\xi$ in (\ref{Espet}) appears to be
practically independent on the particular form of $\chi(x)$ and
corresponds to the Kolmogorov's value $-{5\over 3}$. The  constant $C_K$ is
stable for very narrow $\chi$ functions and approaches the value
$C_K\approx 1.78$ compatible with the experimental results \cite{PO}. These
results are summarized, for 
the energy spectrum $E(q; m)$, in Fig.(\ref{spettr}).

We conclude this section giving a numerical justification of the loop expansion
here considered. A rather rough evaluation in Appendix B shows that the two
and one loop terms 
\be
\label{2loop}
\int {d\hat{q}\over (2\pi)^4}m\partial_mh(q; m)
\Delta_{\hat{v}^{\gamma}v^{\rho}}(\hat{q}; m)
\Gamma_{\tilde{v}^{\rho}(\hat{q})\tilde{v}^{\delta}(-\hat{q})\tilde{v}^{\alpha}
(\hat{p})\hat{v}^{\beta}(0)}\Delta_{v^{\delta}\hat{v}^{\gamma}}(\hat{q}; m),
\ee
\be
\label{1loop}
\int {d\hat{q}\over (2\pi)^4}m\partial_mh(q; m)
\Delta_{\hat{v}^{\gamma}v^{\rho}}(\hat{q}; m)
\Gamma^{(0)}_{\tilde{v}^{\rho}(\hat{q})\tilde{v}^{\alpha}(\hat{p})\hat{v}^s(0)}
\Delta_{\hat{v}^sv^r}(\hat{q} + \hat{p})
\Gamma^{(0)}_{\tilde{v}^r(\hat{q} + \hat{p})
\tilde{v}^{\delta}(-\hat{q})\hat{v}^{\beta}(0)}
\Delta_{v^{\delta}\hat{v}^{\gamma}}(\hat{q}; m)
\ee
are in the ratio 
${1\over n}$ i.e. proportional to the spectrum size of the noise
function $h(x; m)$.
Analogous results are obtained for the other two loops terms and in general the
terms of our loops ordering, generated 
by the ERG, are decreasing as inverse powers
of n.
This result is not a 
sufficient argument  for the validity of the considered loops
expansion, but only a proof of consistency of the theory 
and a precise numerical estimation of all interesting quantities is very
hard. We also observe that, unlike a naive perturbative loops
expansion, the terms of the expansion are not increasing as a power of the
Reynolds number $\cal{R}$.

\newsection{Conclusions}

\noindent
The numerical results in 
the previous section are encouraging and, as reported in
appendix B, an essential ingredient appears to be the size of the spectra of two
point stirring force correlation function. If the approximation scheme here
proposed is physically meaningful, 
the non-perturbative contributions are mainly
due, in the inertial region, to 
the two point function $\Gamma^m_{u\hat{u}}$ and,
particularly, to the anomalous dimension of the $\hat{u}$ field (no anomalous
dimension is associated to the $u$ field as results by the Ward Identity). In
other words, near a critical point, the ERG provides a summability criterion
concerning the two point function and the residual non classical dynamics is
described by perturbative corrections. We cannot provide a proof for this
scenario, but only support it with plausibility arguments since a direct check
requires a very heavy 
numerical analysis. We think that an interesting alternative
to the numerical analysis is 
given by the method recently proposed by T. R. Morris
\cite{TM3}, \cite{TM2} in the context of quantum field theory. The Morris's
method is similar to our analysis in the treatment of the field
anomalous dimensions in 
the ERG, but our perturbative loops expansion is replaced
by a more promising nonperturbative approximation of the effective action. The
Morris's scheme correspond 
to a quasi-local approximation of the effective action
obtained by a derivative expansion. However the
application of this method to our model is not straightforward, in particular
Morris approach does not preserve Galilei invariance. It should however be
interesting to compare numerical results obtained with different methods in the
effort towards a better understanding of the different approximations in the
description of the physics of these systems.

\bigskip
\noindent
{\bf Acknowledgements}

\medskip
\noindent
We wish to thank A. Traverso for pointing out to us the existing literature on
the subject, C. Becchi and G. Gallavotti for helpful discussions
and A. Blasi for a critical revision of the manuscript. One of us (P.T.) is
deeply indebted to G. Curci 
for the hospitality received on the I.N.F.N. APE group
in Pisa.

\renewcommand{\thesection}{A}
\renewcommand{\thesubsection}{A.\arabic{subsection}}

\vspace{12mm}
\pagebreak[3]
\setcounter{section}{1}
\setcounter{equation}{0}
\setcounter{subsection}{0}
\setcounter{footnote}{0}

\begin{flushleft}
{\bf Appendix A}
\end{flushleft}

\noindent
In this appendix we prove that the vertex correlation functions containing only
$v^{\alpha}$ fields necessarily vanish for an arbitrary value of the parameter
$m\not= 0$ if these function vanish for a given $m=\bar{m}$. Consider the flow
equation for any such function; from (\ref{evolut}) we have  
\ba
&m\partial_m\Gamma^m_{\tilde{v}^{\alpha_1}(\hat{p}_1)...v^{\alpha_n}(0)}
\nonumber\\ 
&= -{i\over 2}\int {d\hat{q}\over
\left(2\pi\right)^4}m\partial_mR_{ab}(q; m) \Delta_{\Psi_b\Psi_c}(q; m)
\left.{\delta^{2+n}\bar{\Gamma}(\Psi; m)\over
\delta\tilde{\Psi}_c(\hat{q})\delta\tilde{\Psi}_l(-\hat{q})\delta
\tilde{v}^{\alpha_1}(\hat{p}_1)...\delta v^{\alpha_n}(0)}\right|_{\Psi=0}
\Delta_{\Psi_l\Psi_a}(q; m).\nonumber \\
\label{a1}  
\ea
It is straightforward to see that all terms in the r. h. s. of eq.(\ref{a1})
contain at least one vertex with only the $v^{\alpha}$ field or the
component $\Delta_{v^{\alpha}v^{\beta}}(q; m)$ of the full propagator. For this
reason
$$\left.\partial_m\Gamma^m_{\tilde{v}^{\alpha_1}(\hat{p}_1)...v^{\alpha_n}(0)}
\right|_{m=\bar{m}} = 0$$
and at the generic point $\bar{m} + \delta m$ near $\bar{m}$ we have
\be
\Gamma^{\bar{m}+\delta m}_{\tilde{v}^{\alpha_1}(\hat{p}_1)...v^{\alpha_n}(0)}
= O\left((\delta m)^2\right)
\ee
which iteratively yields the desired result. In other words we have a
non-renormalization property for these vertices and they
will be absent if they are not present in the original {\it classical} action.

\renewcommand{\thesection}{B}
\renewcommand{\thesubsection}{B.\arabic{subsection}}

\vspace{12mm}
\pagebreak[3]
\setcounter{section}{1}
\setcounter{equation}{0}
\setcounter{subsection}{0}
\setcounter{footnote}{0}

\begin{flushleft}
{\bf Appendix B}
\end{flushleft}

\noindent
Here we propose a heuristic argument in order to compare two loops with one
loop terms like the ones considered at the end of section 6. 
Taking into account the results of section 6, consider the two point functions
\be
\Gamma_{\tilde{v}^{\alpha}(\hat{p})\hat{v}^{\beta}(0)}\rightarrow
\Gamma^m_{\tilde{v}^{\alpha}(\tau, x)\hat{v}^{\beta}(0)}\sim 
m^2\left(-i\tau + \sigma\left({D_0\over m^4}\right)^{1\over 3}x^{1+\omega}
\right)P^{\alpha\beta}(n_p).
\label{b2}
\ee
and
\be
\Gamma_{\tilde{\hat{v}}^{\alpha}(\hat{p})\hat{v}^{\beta}(0)}\rightarrow
\Gamma^m_{\tilde{\hat{v}}^{\alpha}(\tau, x)\hat{v}^{\beta}(0)}\sim 
2i{D_0\over m^3}\left(\chi(x) + \gamma x^{-3+\beta}
\right)P^{\alpha\beta}(n_p)
\label{b2m} 
\ee
where, from Table (1), $\omega = 1+\left(-{4\over 3} + \alpha\right)\sim 0.07$
and $\beta\sim 0.4$. The calculation of the vertex function
$\Gamma_{\tilde{v}^{\rho}(\hat{q})\tilde{v}^{\delta}(-\hat{q})\tilde{v}^{\alpha}
(\hat{p})\hat{v}^{\beta}(0)}$ is performed by the ERG where, as in the case of
the two point function, only the leading vertex is taken into account. In this
approximation the ERG which we consider is
\ba
&m\partial_m
\Gamma_{\tilde{v}^{\rho}(\hat{q})\tilde{v}^{\delta}(-\hat{q})\tilde{v}^{\alpha}
(\hat{p})\hat{v}^{\beta}(0)}
\nonumber \\
&= - 2\int {d\hat{k}\over \left(2\pi\right)^4}m\partial_m h(k; m)
\Delta_{\hat{u}^fu^b}(\hat{k}; m)\left\{
\Gamma^{(0)}_{u^b(\hat{k})\hat{u}^{\beta}(-\hat{p})u^s(\hat{k}-\hat{p})}
\Delta_{u^s\hat{u}^l}(\hat{k}-\hat{p}; m)\right.\nonumber \\
&\left(\Gamma^{(0)}_{\hat{u}^l(\hat{k}-\hat{p})u^{\delta}(-\hat{q})
u^m(-\hat{k}+\hat{p}+\hat{q})}\Delta_{u^m\hat{u}^n}(\hat{k}-\hat{p}-\hat{q}; m)
\Gamma^{(0)}_{\hat{u}^n(\hat{k}-\hat{p}-\hat{q})u^{\rho}(\hat{q})
u^a(-\hat{k}+\hat{p})}\right. \nonumber \\
&\left.+ \Gamma^{(0)}_{\hat{u}^l(\hat{k}-\hat{p})u^{\rho}(\hat{q})
u^m(-\hat{k}+\hat{p}-\hat{q})}\Delta_{u^m\hat{u}^n}(\hat{k}-\hat{p}+\hat{q}; m)
\Gamma^{(0)}_{\hat{u}^n(\hat{k}-\hat{p}+\hat{q})u^{\delta}(-\hat{q})
u^a(-\hat{k}+\hat{p})}\right)\nonumber\\
&\Delta_{u^a\hat{u}^d}(\hat{k}-\hat{p}; m)
\Gamma^{(0)}_{\hat{u}^d(\hat{k}-\hat{p})u^{\alpha}(\hat{p})u^c(-\hat{k})} +
\left[\Gamma^{(0)}_{u^b(\hat{k})\hat{u}^{\beta}(-\hat{p})u^s(\hat{k}-\hat{p})}
\Delta_{u^s\hat{u}^l}(\hat{k}-\hat{p}; m)\right.\nonumber \\
&\left(\Gamma^{(0)}_{\hat{u}^l(\hat{k}-\hat{p})u^{\delta}(-\hat{q})
u^m(-\hat{k}+\hat{p}+\hat{q})}\Delta_{u^m\hat{u}^n}(\hat{k}-\hat{p}-\hat{q}; m)
\Gamma^{(0)}_{\hat{u}^n(\hat{k}-\hat{p}-\hat{q})u^{\alpha}(\hat{p})
u^a(-\hat{k}+\hat{q})}\right.\nonumber \\
&\Delta_{u^a\hat{u}^d}(\hat{k}-\hat{q}; m)
\Gamma^{(0)}_{\hat{u}^d(\hat{k}-\hat{q})u^{\rho}(\hat{q})u^c(-\hat{k})} +
\Gamma^{(0)}_{\hat{u}^l(\hat{k}-\hat{p}-\hat{q})u^{\rho}(\hat{q})
u^m(-\hat{k}+\hat{p})}\Delta_{u^m\hat{u}^n}(\hat{k}-\hat{p}
+ \hat{q}; m)\nonumber \\
&\left.\left.\Gamma^{(0)}_{\hat{u}^n(\hat{k}-\hat{p}+\hat{q})u^{\alpha}(\hat{p})
u^a(-\hat{k}-\hat{q})}\Delta_{u^a\hat{u}^d}(\hat{k}+\hat{q}; m)
\Gamma^{(0)}_{\hat{u}^d(\hat{k}+\hat{q})u^{\delta}(-\hat{q})u^c(-\hat{k})}
\right) + \alpha\leftrightarrow\beta\right]\nonumber \\ 
&+ \left[
\Gamma^{(0)}_{u^b(\hat{k})\hat{u}^{\beta}(-\hat{p})u^s(\hat{k}-\hat{p})}
\Delta_{u^s\hat{u}^l}(\hat{k}-\hat{p}; m)
\Gamma^{(0)}_{\hat{u}^l(\hat{k}-\hat{p})u^{\alpha}(\hat{p})
u^m(-\hat{k})}\Delta_{u^m\hat{u}^n}(\hat{k}; m)\right.\nonumber \\
&\left(\Gamma^{(0)}_{\hat{u}^n(\hat{k})u^{\rho}(\hat{q})
u^a(-\hat{k}-\hat{q})}\Delta_{u^a\hat{u}^d}(\hat{k}+\hat{q}; m)
\Gamma^{(0)}_{\hat{u}^d(\hat{k}+\hat{q})u^{\delta}(-\hat{q})u^c(-\hat{k})}
+\right.\nonumber \\
&\left.\left(\Gamma^{(0)}_{\hat{u}^n(\hat{k})u^{\delta}(-\hat{q})
u^a(-\hat{k}+\hat{q})}\Delta_{u^a\hat{u}^d}(\hat{k}-\hat{q}; m)
\Gamma^{(0)}_{\hat{u}^d(\hat{k}-\hat{q})u^{\rho}(\hat{q})u^c(-\hat{k})}
\right) + \alpha\leftrightarrow\beta\right]\nonumber\\
&+\left(
\Gamma^{(0)}_{u^b(\hat{k})u^{\delta}(-\hat{q})\hat{u}^s(\hat{k}-\hat{q})}
\Delta_{\hat{u}^su^l}(\hat{k}-\hat{q}; m)
\Gamma^{(0)}_{u^l(\hat{k}-\hat{q})\hat{u}^{\beta}(-\hat{p})
u^m(-\hat{k}+\hat{q}+\hat{p})}\Delta_{u^m\hat{u}^n}(\hat{k}-\hat{q}-\hat{p}; m)
\right.\nonumber \\
&\left.\left.\Gamma^{(0)}_{\hat{u}^n(\hat{k}-\hat{q}-\hat{p})u^{\alpha}(\hat{p})
u^a(-\hat{k}+\hat{q})}\Delta_{u^a\hat{u}^d}(\hat{k}-\hat{q}; m)
\Gamma^{(0)}_{\hat{u}^d(\hat{k}-\hat{q})u^{\rho}(\hat{q})u^c(-\hat{k})}
+ \delta\leftrightarrow\rho\right)\right\}\Delta_{u^c\hat{u}^f}(\hat{k}; m).
\label{RG4}  
\ea
We rescale the variables $(0, p^{\alpha})$, $(q^0, q^{\rho})$ and $(k^0,
k^{\delta})$ as
\be
\label{riscala}
p^{\alpha} = mxn^{\alpha}, \quad q^0 = m^2z^0, \quad q^{\rho} = mzn^{\rho},
\quad k^0 = m^2y^0, \quad k^{\delta} = myn^{\delta}
\ee
where $n^{\alpha}$, $n^{\rho}$ and $n^{\delta}$ are unitary vectors along
the directions 
of $p^{\alpha}$,  $q^{\rho}$ and $k^{\delta}$. We also observe that
the value of 
$\left|\vec{k}\right|$ is constrained, by the support of the function
$h(k; m)$ (see (\ref{HN}) and (\ref{noispec})), into a narrow region centered
around $k\sim m$, so that, from (\ref{riscala}), $y\sim 1$. Moreover, in order
to evaluate the two loops term (\ref{2loop}), the variable $z$ has also the same
constraint. From the last observation it follows that, in the inertial region
(where $x\gg 1$), we can consider ${y\over x}\sim {z\over x}\ll 1$.
Taking into account only the leading terms we obtain from eq.(\ref{RG4}), after
tedious but straightforward calculations
\ba
&\left(m\partial_m - x\partial_x - z\partial_z - 2z^0\partial_{z^0}\right)
\Gamma_{\tilde{v}^{\rho}(\hat{q})\tilde{v}^{\delta}(-\hat{q})\tilde{v}^{\alpha}
(\hat{p})\hat{v}^{\beta}(0)}
\nonumber \\
&\sim {1\over 6\pi^2}{1+\omega\over n}\Gamma\left({2+s-\omega\over 2n}\right)
\left({D_0\over m^4}\right)^{1\over
3}\left[{x^{1-\omega}n_p^sn_p^{\rho}P^{\alpha\beta}(n_p)\over
\sigma^2\left((z^{0})^{2}+ \sigma^2\left({D_0\over m^4}\right)^{2\over 3}
x^{2+2\omega}\right)} + ...\right] + \int dy\chi(y)O\left({y\over x}, {z\over
x}\right).
\label{particol}  
\ea
Where we have considered $\chi(y) = y^se^{-y^{2n}}$ as in (\ref{noispec}). A
solution of eq.(\ref{particol}) is not obtainable with elementary procedures and
requires numerical methods, but it is still possible to extract sufficient
information concerning our purpose with very simple arguments. We simplify
(\ref{particol}) rescaling the variable $z^0$ as
\be
\label{zrisc} 
z^0 = \sigma\left({D_0\over m^4}\right)^{1\over 3}x^{1+\omega}w^0
\ee
and setting
\be
\label{soluz}
\Gamma_{\tilde{v}^{\rho}(\hat{q})\tilde{v}^{\delta}(-\hat{q})\tilde{v}^{\alpha}
(\hat{p})\hat{v}^{\beta}(0)}
= {1\over 6\pi^2}{1+\omega\over n}\Gamma\left({2+s-\omega\over 2n}\right)
\left({D_0\over m^4}\right)^{-{1\over 3}}{x^{-(1+3\omega)}\over \sigma^4}
G(m, x, w^0)n_p^{\delta}n_p^{\rho}P^{\alpha\beta}(n_p) + ...
\ee
The function $G(m, x, w^0)$ satisfies the equation

$$\left(m\partial_m - x\partial_x - 2w^0\partial_{w^0}\right)G(m, x, w^0)
- \left({1\over 3}-3\omega\right)G(m, x, w^0) = {1\over 1 + (w^0)^2}.$$

\noindent
Inserting (\ref{soluz}) in (\ref{2loop}) we obtain
\ba
&\int {d\hat{q}\over (2\pi)^4}m\partial_mh(q; m)
\Delta_{\hat{v}^{\gamma}v^{\rho}}(\hat{q}; m)
\Gamma_{\tilde{v}^{\rho}(\hat{q})\tilde{v}^{\delta}(-\hat{q})\tilde{v}^{\alpha}
(\hat{p})\hat{v}^{\beta}(0)}\Delta_{v^{\delta}\hat{v}^{\gamma}}(\hat{q}; m)
\nonumber\\
&= {m^2\over 12\pi^4}{1+\omega\over n}\Gamma\left({2+s-\omega\over 2n}\right)
\left({D_0\over m^4}\right)^{1\over 3}{x^{-2\omega}\over \sigma^5}
P^{\alpha\beta}(n_p)\int dzz^{-2\omega}\left(3 +s
-2nz^{2n}\right)z^se^{-z^{2n}}G\left(m, x, {z\over x}\right).\nonumber\\
&+ .............\nonumber
\ea
Considering the leading term $\left(G\left(m, x, {z\over x}\right)
= G^{\prime}\left(m, x\right) + O \left({z\over x}\right)\right)$ we have
\ba
&\int {d\hat{q}\over (2\pi)^4}m\partial_mh(q; m)
\Delta_{\hat{v}^{\gamma}v^{\rho}}(\hat{q}; m)
\Gamma_{\tilde{v}^{\rho}(\hat{q})\tilde{v}^{\delta}(-\hat{q})\tilde{v}^{\alpha}
(\hat{p})\hat{v}^{\beta}(0)}\Delta_{v^{\delta}\hat{v}^{\gamma}}(\hat{q}; m)
\nonumber\\
&= {m^2\over 12\pi^4}\left({1+\omega\over 
n}\right)^2\Gamma\left({2+s-\omega\over
2n}\right)\Gamma\left({1+s-2\omega\over
2n}\right) \left({D_0\over m^4}\right)^{1\over 3}{x^{-2\omega}\over \sigma^5}
P^{\alpha\beta}(n_p)G^{\prime}\left(m, x\right) + ....\nonumber
\ea
The Euler's functions $\Gamma\left({2+s-\omega\over 2n}\right)$ and
$\Gamma\left({1+s-2\omega\over 2n}\right)$ are of order of unity.
With the same method we prove that (\ref{1loop}) is order
$O\left({1\over n}\right)$. The previous arguments show a decreasing with
the powers of the spectrum size of the noise function $h(x; m)$ for the
considered loops ordering. This argument gives a meaning to the last term of
Tab:2 in Section 6. Finally we observe that the previous discussion is not a
proof, but only a consistency argument with the hypothesis of Section 5.

\end{document}